\documentclass[12pt, a4paper]{article}
\usepackage{jheppub}

\usepackage{amsmath,amssymb}
\usepackage{xcolor}
\usepackage{comment}
\usepackage{setspace}

\newcommand{\vk}{{\vec{k}}}
\newcommand{\vx}{{\vec{x}}}
\newcommand{\vy}{{\vec{y}}}
\newcommand{\vz}{{\vec{z}}}
\newcommand{\cA}{{\cal A}}

\newcommand{\cK}{{\cal K}}
\newcommand{\cL}{{\cal L}}
\newcommand{\cM}{{\cal M}}

\newcommand{\tf}{{\tilde f}}

\newcommand{\tH}{{\tilde H}}
\newcommand{\tR}{{\tilde R}}
\newcommand{\tphi}{{\tilde\phi}}
\newcommand{\eps}{\epsilon}

\newcommand{\be}{\begin{eqnarray}}
\newcommand{\ee}{\end{eqnarray}}

\newcommand{\nt}{\notag\\}

\newcommand{\pfrac}[2]{\frac{\partial{#1}}{\partial{#2}}}

\preprint{KEK-TH-2297}

\title{Calculation of Hawking Radiation in Local Field Theory}

\author[a]{Shotaro Shiba Funai}
\author[b]{and Hirotaka Sugawara}

\affiliation[a]{Physics and Biology Unit, Okinawa Institute of Science and Technology (OIST),\\1919-1 Tancha Onna-son, Kunigami-gun, Okinawa 904-0495, Japan}
\affiliation[b]{High Energy Accelerator Research Organization (KEK),\\1-1 Oho, Tsukuba, Ibaraki 305-0801, Japan}

\emailAdd{shotaro.funai@oist.jp}
\emailAdd{sugawara@post.kek.jp}

\abstract{
Hawking radiation~\cite{ref1,ref1a} of {the} blackhole~\cite{ref2} is calculated 
based on the principle of local field theory. % principle.
In our approach, the radiation is a unitary process, therefore no information loss will be
recorded. In fact, 
observers in different regions of the space communicate using the Hawking radiation,
when the systems in the different regions are entangled with each other.
The entanglement entropy of the blackhole is also calculated in the local field theory. 
{We} %It is 
found that the entanglement entropy of the systems separated
by the {blackhole horizon} % of a blackhole 
is closely connected to the Hawking radiation in our approach.
Our calculation shows that  
the entanglement entropy of the systems separated by the horizon of a blackhole is just a pure number
$\frac{\pi^3 + 270 \zeta(3)}{360 \pi^2}$,
independent of any parameter of the blackhole,
and its relation to the Hawking radiation is given by
$S_{EE} = %{\frac{1}{\sqrt 2}} 
\frac{8 \pi}{3} \frac{\pi^3 + 270 \zeta(3)}{\pi^3 + 240 \zeta(3)}
\cA R_H$, 
where $S_{EE}$ is the entanglement entropy, $\cA$ is the area of the horizon, and $R_H$ is the Hawking radiation.}

\begin{document}

\maketitle

%%%%%%%%%%%%%%%%%%
\section{Introduction}
%%%%%%%%%%%%%%%%%%

The information loss {occurs} due to the transition from unitary process to thermodynamic or
stochastic process. In this paper, we show that it will not happen in the case of blackhole. 
This % It
means that the blackhole solution corresponds to a pure but an entangled state.
%, although the real blackhole can be more complicated as will be discussed in Sec.\,\ref{sec:5} of this paper.

Consider a system with Hamiltonian $H$ which is the sum of two parts $H_A, H_B$:
\be
H = H_A + H_B
\ee
where $A$ and $B$ are % can be 
different parts of space or %can be 
composed of different particles.
If $H_A$ and $H_B$ commute with each other, i.e., $[H_A, H_B] = 0$,
the two parts are independent and the energy eigenstate of the whole system can be written as
\be
|\psi\rangle = |\psi_A\rangle |\psi_B\rangle
\ee
where $|\psi_A\rangle$ and $|\psi_B\rangle$ are energy eigenstates of $H_A$ and $H_B$,
respectively.
However, % But 
if $H_A$ and $H_B$ do not commute with each other, we have
\be
|\psi\rangle = |\psi_A\rangle |\psi_B\rangle + |\psi'_A\rangle |\psi'_B\rangle + \cdots
\ee
where $|\psi'_A\rangle$ and $|\psi'_B\rangle$ are other eigenstates. 
The state $|\psi\rangle$ is an entangled state in this case.
To describe such a situation, we have to take into account that both $H_A$ and $H_B$ are time-dependent, % by fixing certain time. 
{then we find} %It is then clear 
that there is certain energy flow between the systems $A$ and $B$.

The blackhole is such a pure but entangled state as will be shown in this paper: 
$A$ in this case corresponds to {inside} of the blackhole horizon and 
$B$ is the {outside}. % of the horizon.
$H_A$ and $H_B$ do not commute due to the existence of the boundary. 
There must be an energy flow through the boundary, {which} %and it 
is in the form of Hawking radiation. % in the blackhole case.

We calculate this energy flow based on the local field theory and obtain the result which is
similar to the Planck formula but not quite:
This result does not allow the conventional interpretation of the blackhole system to be a mixed state characterized by certain temperature.
We also calculate the entanglement entropy of the blackhole and show that it is related to the Hawking radiation.
% as is pointed out in the abstract. 

{This paper is organized as follows.}
In Sec.\,\ref{sec:2}, we explain our formulation of the local field theory
which is characterized by Schwinger commutation relation among the energy
momentum tensor components.
This gives a formula for the energy flow through the boundary of the two regions $A$ and $B$.
The Schwinger commutation relation cannot be applied to the quantum spin-2 particle.
Then in Sec.\,\ref{sec:3}, we treat the gravity as classical so that we can safely use it.
In Sec.\,\ref{sec:4}, we calculate the energy flow just outside of the horizon
%. For this purpose, it is convenient to use 
using the Rindler coordinate, which is free of singularity at the horizon.
%The Rindler time variable $\tau$ is common to both inside and outside of the horizon,
%and the {radial} variable $\rho$ is real positive outside %of
%the horizon and negative imaginary inside the horizon.
We also calculate the entanglement entropy $S_{EE}$ of the blackhole 
and show that it is a pure number 
$\frac{\pi^3 + 270 \zeta(3)}{360 \pi^2}$
related to the Hawking radiation $R_H$ as
\be
S_{EE} = %\frac{1}{\sqrt 2}
\frac{8 \pi}{3} \frac{\pi^3 + 270 \zeta(3)}{\pi^3 + 240 \zeta(3)}
%\frac{2 \pi^3}{3} \frac{\pi^3 + 270 \zeta(3)}{\pi^5 + 45 \zeta(5)}
\cA R_H
\ee
%is pointed out in the abstract. 
where $\cA$ is the area of the horizon.
\section{Formulation}
\label{sec:2}
%%%%%%%%%%%%%%%%%%

\subsection{Schwinger commutation relation}

We consider four-dimensional Lorentz-invariant local field theory. 
Schwinger proved in 1963~\cite{ref4} that in such a case %we generally have
the energy-momentum tensor $\Theta_{\mu\nu}$ generally satisfies
\be\label{eq1}
[\Theta_{00}(x), \Theta_{00}(y)] = -i\left( \Theta_{0i}(x)+\Theta_{0i}(y) \right)
\partial_i\delta(\vx-\vy)
\ee
where the index $i=1,2,3$ and $\vx, \vy$ are spatial vectors.
{The case with the quantum spin-2 field} % The quantum spin 2 case 
must be excluded from this formula,
and the case of classical gravitational %Schwarzschild 
background~\cite{ref2} will be discussed 
{in the following sections}. %shortly.

If we divide the space into two regions $A$ and $B$ with %the help of 
the functions $f_A$ and $f_B$:
\be\label{eq1a}
f_A=\begin{cases}
1 & \text{in $A$ and on the boundary of $A$ and $B$} \\
0 & \text{otherwise}
\end{cases}\,,\qquad
f_B=\begin{cases}
1 & \text{in $B$}\\
0 & \text{otherwise}
\end{cases}\,,
\ee
meaning that we assume the boundary belongs to $A$ rather than $B$.
Then we define the Hamiltonians $H_A, H_B$ as
\be\label{eq2}
[H_A, H_B] 
&:=& 
\left[ \int_A dx\, f_A(x) \Theta_{00}(x) , \int_B dy\, f_B(y) \Theta_{00}(y) \right] \nt
%&=& -i \int_B dy \int_A dx\, f_A(x) f_B(y) \left\{\Theta_{0M}(x) + \Theta_{0M}(y)\right\}
% \partial_M \delta(\vx-\vy) \nt
%&=& i \int_{B\cap A} dy \left\{
%f_B\partial_M \left(f_A\Theta_{0M}(y)\right) - f_A\partial_M \left(f_B\Theta_{0M}(y)\right) \right\}
%\nt &&
%+\, i \int_B dy \int_{\partial A} dx\, f_A(x) f_B(y) \Theta_{0n}(x) \delta(\vx-\vy)
%\nt &&
%+\, i \int_{\partial B} dy \int_A dx\, f_A(x) f_B(y) \Theta_{0n}(x) \delta(\vx-\vy) \nt
&=& i \int_{A\cap B} dy\, \left(f_B\partial_i f_A - f_A \partial_i f_B\right) \Theta_{0i}(y) \nt
&{=}& -i \int_{bdy} ds\, \Theta_{0n}(s)\,.
\ee

The final form is {obtained since} %when 
$A\cap B$ collapses to the boundary of $A$ and $B$. 
This formula simply means that we have energy-momentum flow between the regions $A$ and $B$.
Here $n$ is the direction normal to the boundary (to the outside direction of $A$) 
and $ds$ is the surface area taken to be the normal vector to the surface.

%{In the second to the last form,} the integration is on the surface of $B\cap A$. 
{This formula also means that} % This means that 
Hamiltonians $H_A, H_B$ for the regions $A$ and $B$ do not generally commute unless the common region $A\cap B$ vanishes, i.e., $A\cap B=\emptyset$.

\subsection{Boundary}
A system described by Hamiltonian 
\be
H &=& \int_{{A\cup B}} \Theta_{00}(x) dx 
= \int_A f_A(x)\Theta_{00}(x) dx + \int_B f_B(x)\Theta_{00}(x) dx \nt
&=& H_A+H_B
\ee
cannot be described as the two independent systems described by $H_A$ and $H_B$, respectively, because $H_A$ and $H_B$ do not commute as shown above. 
The time dependence of $H_A$ and $H_B$ can be calculated as % follows:
\be
i \partial_t H_A &=& [H, H_A] = [H_B, H_A] = i \int_{bdy} \Theta_{0n}(s) ds\,, \nt
i \partial_t H_B &=& [H, H_B] = [H_A, H_B] = -i \int_{bdy} \Theta_{0n}(s) ds\,.
\ee
Therefore, we obtain
\be
H_A(t) &=& H_A(0) + \int_0^t dt \int_{bdy} \Theta_{0n}(s) ds\,,\nt
H_B(t) &=& H_B(0) - \int_0^t dt \int_{bdy} \Theta_{0n}(s) ds\,.
\ee

\subsection{Boundary Hamiltonian}
The next question is
if we can separate out the boundary energy from $H_A$ in such a way that
\be
&&
[\tH_A, \tH_B] = 0\,, \nt
&&
[\tH_{bdy}, \tH_A] = -[\tH_{bdy}, \tH_B] = i \int_{bdy} \Theta_{0n}(s) ds\,.
\ee
We may write 
\be \label{tildeH}
\tH_A = \int dx\, (1 - {\tf_A}(x)) \Theta_{00}(x)\,,\quad
\tH_B = \int dx\, (1 - {\tf_B}(x)) \Theta_{00}(x)\,,
\ee
where the functions $\tf_A, \tf_B$ are %different from Eq.\,(\ref{eq1a}) but 
defined to be 
% is defined to be (we use the same notation $f$ and $g$ here as before but completely different): \ref{eq1a}
\be
\begin{cases}
\tf_A(x)=1 & \text{in $B$} \\
0<\tf_A(x)<1 & \text{in $A$ and close to the boundary with $B$} \\
\tf_A(x)=0 & \text{in $A$ and away from the boundary}
\end{cases}
\ee
and
\be
\begin{cases}
\tf_B(x)=1 & \text{in $A$}\\
0<\tf_B(x)<1 & \text{in $B$ and close to the boundary with $A$} \\
\tf_B(x)=0 & \text{in $B$ and away from the boundary}
\end{cases}\,.
\ee
Then we have
\be
H = \tH_A + \tH_B + \tH_{bdy}
\ee
with
\be
\tH_{bdy} = \int_A dx\, \tf_A(x) \Theta_{00}(x) + \int_B dx\, \tf_B(x) \Theta_{00}(x)
=: \int dx\, \zeta(x) \Theta_{00}(x)\,.
\ee
where
$\zeta(x) = 1$ at the boundary %($r = R$ when rotationally symmetric) 
and falls off rapidly away from the boundary.

{Note that this} % Our 
$\tH_{bdy}$ obviously has nothing to do with the conjectured AdS/CFT correspondence~\cite{ref5}
where a specific {gravity} solution in the bulk corresponds to a specific 
{field theory} % model 
at the boundary. 
We are merely taking the limit of an arbitrary field configuration in local field theory to the boundary, excluding the possibility that spacetime itself has a boundary:
we have {always both} inside and outside of the boundary.

% We may take 
In the case of rotationally symmetric {spacetime, for example,}
\be \label{eq13}
\tf_A(r) = \theta(r - (R - \eps))\,,\qquad
\tf_B(r) = \theta((R + \eps) - r)
\ee
where {$\theta(x)$ is Heaviside step function, then}
the boundary is at $r=R$, and $\eps \ll R$.
{This means the region of $A$ is inside of the boundary, while $B$ is outside.}
We then have
\be
[\tH_A,\tH_B] = 0
\ee
and
\be 
{}[\tH_A,\tH_{bdy}] 
%&=&
%\left[ \int dx\, (1-\tf_A(x))\Theta_{00}(x), 
%\int dx\, \zeta(x)\Theta_{00}(x)\right] \nt
&=&
-i \int_{bdy} ds\,\Theta_{0r}\Big|_{r=R-\eps} \,, \nt
{}[\tH_B,\tH_{bdy}] 
&=&
i \int_{bdy} ds\,\Theta_{0r}\Big|_{r=R+\eps} \,.
\ee

\subsection{Non-entangled pure states}

In the ordinary quantum theory, if the ground state is a non-entangled pure state
$|0\rangle$, it satisfies
\be
\langle 0|[H_A, H_B]|0\rangle
= \langle 0|i\int_{bdy} \Theta_{0n}(s)ds|0\rangle
= 0\,.
\ee
This allows us to ignore the non-commutativity of $H_A$ and $H_B$ at the boundary
when they act on $|0\rangle$,
making %and {this} makes 
it possible to write
\be
|0\rangle = |0\rangle_A \otimes |0\rangle_B\,.
\ee
Notable exceptions are the following:
\begin{enumerate}
\item
In two-dimensional conformal theory where the commutator
$[\Theta_{00}(x) , \Theta_{00}(y)]$ has anomaly~\cite{ref6},
its vacuum expectation value does not vanish. 
This leads to an entangled vacuum state.
\item
When we deal with the quantum theory with the classical blackhole background ground
state $|\Omega\rangle$, we can show that
\be
\langle \Omega|i\int_{bdy} \Theta_{0n}(s)ds|\Omega\rangle
\neq 0\,.
\ee
In fact, this provides the amount of radiation from a hidden region $A$ or $B$,
i.e., Hawking radiation.
We will discuss the case of blackholes in detail below.
\end{enumerate}

%%%%%%%%%%%%%%%%%%%%%%%%%%%%%%%%%%%%%%%%%%%%%%%%
\section{Application to the system with classical gravity}
\label{sec:3}
%%%%%%%%%%%%%%%%%%%%%%%%%%%%%%%%%%%%%%%%%%%%%%%%

\subsection{Formalism}

We apply the above formulation to the neutral scalar system with gravity background, especially with the blackhole background~\cite{ref2}.
Here we are interested in the system defined by the %following 
Lagrangian
\be \label{eq:Lag}
L = \int d^4x\, \cL\,; \quad 
\cL = -\sqrt{{-g}} \left( g^{\mu\nu}\partial_\mu \phi \partial_\nu \phi + M^2 \phi^2 \right).
\ee
%\funai{$g$ isn't negative? Is the spacetime Euclideanized?} 
Then the energy-momentum tensor is obtained as
\be
\Theta_{\mu\nu} 
= \frac{1}{\sqrt{-g}}\frac{\partial\cL}{\partial g^{\mu\nu}}
= -\partial_\mu \phi \partial_\nu \phi + \frac12 g_{\mu\nu}
\left(\partial^\rho \phi \partial_\rho \phi + M^2 \phi^2 \right).
\ee
Assuming $g_{\mu\nu}$ is diagonal, we obtain
\be
\Theta_{0{i}}(s) = -\partial_0 \phi \partial_i \phi
\ee
and
\be\label{eq23}
\Theta_{00} 
= -\frac12(\partial_0 \phi)^2
 + \frac12 g_{00} \left(\partial^i \phi \partial_i \phi + M^2 \phi^2 \right)
\ee
{where the index $i$ runs only spatial directions $1,2,3$.}
In principle, the classical self-consistent gravity is defined as
\be
R_{\mu\nu} - \frac12 g_{\mu\nu} R
=8\pi G \langle \Theta_{\mu\nu} \rangle + (\text{other sources}),
\ee
but here we treat $g_{\mu\nu}$ as given or just as a solution
to the Einstein equation without scalar fields~\cite{ref2}. Therefore,
the ground state (vacuum state without scalar particles) is described by $g_{\mu\nu}$
(e.g., a blackhole) with certain zero point energy.

We now consider the region $A$ ($B$) to be the inside (outside) of three-dimensional space with spherical symmetry, respectively.
In this case, we have
\be
ds^2 = {g_{00}}(r,t)dt^2 + {g_{rr}}(r,t)dr^2 + r^2(d\theta^2+\sin^2\theta d\varphi^2)
\ee
where
$\sqrt{-g}=\sqrt{-g_{00}g_{rr}}\,r^2\sin\theta$.
%and 
%$\Theta_{0i}=-\partial_0\phi\partial_i\phi$.

\subsection{Commutation relation}

We calculate the commutation relation $[\Theta_{00}(x), \Theta_{00}(y)]$ directly using
\be\label{eq28}
[P_\phi,\phi]
%:= [-g^{00}\partial_0\phi,\phi]
=-g^{00}[\partial_0\phi,\phi]
=i\delta^3(\vx-\vy)
\ee
{where the conjugate momentum
$P_\phi := \frac{1}{\sqrt{-g}}\frac{\partial \cL}{\partial (\partial_0\phi)}$.}
Then, from Eq.\,(\ref{eq23}), %(\ref{\eq27}) and (\ref{eq28}), 
we obtain
\be
[\Theta_{00}(x),\Theta_{00}(y)]
= -ig_{00}(x)g_{00}(y)\left( g^{ij}\Theta_{0j}(x) + g^{ij}\Theta_{0j}(y) \right)
  \partial_i \delta^3(\vx-\vy)\,.
\ee
This is the same as the case of no gravity, i.e., Eq.\,(\ref{eq1}), 
if we replace $\Theta_{0 i}(x)$ in the right hand side with $g^{ij}\Theta_{0 j}(x)$
{and} except for the factor $g_{00}(x) g_{00}(y)$.
Taking into account {these factors}, %the commutation relation (\ref{eq28}),
the Hamiltonian is defined as 
\be
H = \int d^{3}x\, \sqrt{-g}\,g^{00}\Theta_{00}(\vx) \,.
\ee

Then we have, as in the case without gravity background (\ref{eq2}),
\be
[H_A, H_B] 
&=&
\left[\int d^3x\,\sqrt{-g}\,f_A g^{00}\Theta_{00}(\vx),
 \int d^3y\,\sqrt{-g}\,f_B g^{00}\Theta_{00}(\vy)\right] \nt
&=&
{ i\int_{A\cap B} d^3x\,\sqrt{-g} } \left( f_B\partial_i f_A - f_A\partial_i f_B\right)
 g^{ij}\Theta_{0j}(\vx) \,,
\ee
{where $f_A, f_B$ are defined as in Eq.\,(\ref{eq1a}),
meaning that} $f_A = 1$ and $f_B = 0$ at the boundary of $A$ and $B$. % as before.
{Since} %Then, if 
$A\cap B$ is just the boundary of $A$ and $B$ {in our case}, it becomes
\be
[H_A,H_B] 
&{=}& -i\int_{bdy} ds\,\sqrt{-g}\, g^{rr} \Theta_{0r} 
= -i\int_{bdy} ds\, \sqrt{-\frac{g_{00}}{g_{rr}}} r^2\sin\theta\, \Theta_{0r}\,.
\ee

We define $\tH_A, \tH_B$ as in Eq.\,(\ref{tildeH}):
\be
\tH_A &:=& \int d^3x\,\sqrt{-g}\, g^{00}(1-\tf_A(\vx))\Theta_{00}(\vx) \,, \nt
\tH_B &:=& \int d^3x\,\sqrt{-g}\, g^{00}(1-\tf_B(\vx))\Theta_{00}(\vx) \,.
\ee
Then we obtain
\be
[\tH_A,\tH_B]=0
\ee
and
\be
[\tH_A,\tH_{bdy}]
= -i\int_{bdy} ds\sqrt{-\frac{g_{00}}{g_{rr}}} r^2\sin\theta\, \Theta_{0r}\Big|_{r=R-\eps}
\ee
where {$R, \eps$ are defined in Eq.\,(\ref{eq13}),
and $\tH_{bdy}:=H- \tH_A -\tH_B$ is written as} % given by
\be
\tH_{bdy} =:  \int d^3x\,\sqrt{-g}\, \zeta(\vx) g^{00} \Theta_{00}(\vx)\,.
\ee
In the case of Schwarzschild solution, we have
\be
g_{00} = -\left(1-\frac{r_s}{r}\right)\,,\qquad
g_{rr} = \frac{1}{1-\frac{r_s}{r}}\,,
\ee
{with the boundary radius $R=r_s$}, and the boundary Hamiltonian is 
\be
\tH_{bdy}
&=&\frac12 \int_{bdy'} \left[
 \frac{(\partial_0\phi)^2}{1-\frac{r_s}{r}}
 +\left(1-\frac{r_s}{r}\right)(\partial_r\phi)^2
 +\frac{1}{r^2}\left((\partial_\theta\phi)^2 + \frac{(\partial_\varphi \phi)^2}{\sin^2\theta}
 \right) + M^2\phi^2\right] \nt
 &&\qquad\times\,
 r^2\sin\theta\, drd\theta d\varphi % {dt}
\ee
where the integration {region ($bdy'$) is} % done in 
a thin sphere of depth $2\eps$. 
If we take the boundary to be exactly at $r = r_s$, the inside %of $r_s$ 
and the outside have different signs of $(1 - \frac{r_s}{r})$. 
{To avoid this ambiguity,}
we assume the boundary to be in $r \in [r_s - \eps, r_s + \eps]$
{and in this region $\zeta(\vx)=1$ is satisfied.}

%%%%%%%%%%%%%%%%%%%%%%%%%%%%%%%%%%%%%%%%%%%%%%%%%%%%%%%%%
\section{Calculation of Hawking Radiation and entanglement entropy}
\label{sec:4}
%%%%%%%%%%%%%%%%%%%%%%%%%%%%%%%%%%%%%%%%%%%%%%%%%%%%%%%%%

\subsection{Preliminary}

Our definition of the Hawking radiation is
\be\label{eq39}
R_H := \frac{\partial H_B}{\partial t} 
= {-i}\langle \Omega|[H_A,H_B]|\Omega\rangle
= \int_{bdy} ds\sqrt{-\frac{g_{00}}{g_{rr}}}\, r^2\sin\theta
\langle \Omega|\Theta_{0r}|\Omega\rangle
\ee
where $|\Omega\rangle$ is the ground state of 
$H = \tH_A + \tH_B + \tH_{bdy}$.
We sometimes call $|\Omega\rangle$ the Hawking ground state.
$R_H$ is clearly 
the amount of energy going through the boundary in the ``vacuum state'' which is nothing but the blackhole. 
This way of interpreting the Hawking radiation has two remarkable characters:
\begin{enumerate}
\item
The Hawking radiation process is the unitary process. Therefore, no violation of information
conservation occurs. Explicitly, the above equation shows
\be \label{eq40}
%\langle\Omega||\Omega\rangle
\int dt \,\langle\Omega|R_H|\Omega\rangle
= \langle\Omega|H_B(t)|\Omega\rangle
= \langle\Omega|U^\dagger(t)H_B(0)U(t)|\Omega\rangle
\ee
{where $U(t)=e^{iHt}$ is the time evolution operator.}
%\funai{??}
Therefore, we obtain % have
\be \label{eq41}
\langle\Omega|i\frac{\partial U}{\partial t}|\Omega\rangle
= {-}\langle\Omega|UH_A{(t)}|\Omega\rangle.
\ee
%\funai{??}
\item
Since the time development of $H_B$ is completely given by the Hawking radiation, we can say that an observer in $A$ and an observer in $B$ are exchanging the information using the Hawking radiation. 

Extending this situation to other entangled states such as the EPR system~\cite{ref7},
we can say {that} Alice in $A$ and Bob in $B$ are exchanging the information using the Hawking radiation. 
In the case of EPR, there must be a ``Hawking'' radiation emitted to maintain the entanglement nature of $A$ and $B$. 
{An} equation similar to (\ref{eq39}) provides a way to calculate the Hawking radiation
{also} in this case. % also. 
In other words, we are claiming EPR\,$=$\,SH (Steven Hawking) rather than EPR\,$=$\,ER~\cite{ref8}
where the information is exchanged using a wormhole.
\end{enumerate}

\subsection{Rindler coordinate formulation due to 't Hooft} % ~\cite{ref2}}
\label{sec:3.3}

Now we have to calculate $R_H$ in Eq.\,(\ref{eq39}) explicitly. 
It is important to understand rather tricky nature of Eq.\,(\ref{eq39}): 
To calculate the left hand side of Eq.\,(\ref{eq39}), 
we must be careful about what happens at the boundary. We must use
\be
H = \tH_A + \tH_B + \tH_{bdy}
\ee
where $[\tH_A,\tH_B]=0$ and 
\be\label{eq36}
{[\tH_B, \tH_{bdy}]}
= i\int_{bdy} ds \sqrt{-\frac{g_{00}}{g_{rr}}} \, r^2\sin\theta\,\Theta_{0r} \Big|_{r=r_s+\epsilon}\,.
\ee
However, after taking the commutator and {ending} % ended 
up with the expression on the right hand side, 
we may use the boundary value on the right hand side of Eq.\,(\ref{eq39}) 
just {by} extending the following expression to the boundary $r=r_s$,
\be
\tH_A\to H_A\,,\qquad
\tH_B\to H_B\,.
\ee
{This means that the commutator singularity of $[\tH_B, \tH_{bdy}]$ on the boundary is already taken into account on the right hand side of (\ref{eq36})}.
%\funai{it should be explained further.}
On the right hand side, we have
\be
\Theta_{0 r} = -\partial_0\phi \partial_r\phi
\ee
where the scalar field $\phi$ is a free field and can be expanded in normal mode in the entire space, space $A$, or space $B$. 

We basically follow 't Hooft's formulation~\cite{ref2} in the following.
To get the normal mode, we {need to} solve the equation of motion.
This can be done by changing the coordinate system to Rindler coordinate. %~\cite{ref2}.
Since the right hand side of Eq.\,(\ref{eq39})
contains only the quantity at the boundary (i.e., horizon), the transformation 
from $(t,r,\theta,\varphi)$ to the Rindler system $(\tau,\rho,\theta,\varphi)$
becomes very simple:
\be
\tau = \frac{1}{2r_s} t\,,\quad
\rho = 2r_s\sqrt{\frac{r}{r_s}-1}\,,
\ee
and
\be
g_{\tau\tau} = \pfrac{t}{\tau}\pfrac{t}{\tau} g_{00} = -\frac{r_s}{r}\rho^2 \to -\rho^2\,,\qquad
g_{\rho\rho} = \pfrac{r}{\rho}\pfrac{r}{\rho} g_{rr} = \frac{r}{r_s} \to 1\,,
\ee
in the limit of $r\to r_s$ {(i.e., $\eps\to +0$)}.
The equation of motion {for the Lagrangian (\ref{eq:Lag})} becomes
\be \label{eq46}
\left[ \left(\rho\pfrac{}{\rho}\right)^2 - \pfrac{^2}{\tau^2} 
%+ \rho^2\left(\pfrac{^2}{\vz^2}-M^2\right) \right] \phi=0
- \rho^2\left(\frac{\ell(\ell+1)}{r_s^2}+M^2\right) \right] R_\ell(\rho,\tau)=0
\ee
where $\phi(\tau,\rho,\theta,\varphi) =: R_\ell(\tau,\rho)Y_{\ell m}(\theta,\varphi)$
and $Y_{\ell m}$ are the spherical harmonics.
%
\begin{comment}
$\vz := {r_s\times}(\theta,\varphi)|_{r=r_s, \theta\approx\frac{\pi}{2}}$
is defined.
{This means that we discuss only the region with $\sin\theta\approx 1$, satisfying}
%This means we approximate
\be
\frac{1}{r_s^2}\left(
\frac{1}{\sin\theta}\pfrac{}{\theta}(\sin\theta\pfrac{}{\theta})
 + \frac{1}{\sin^2\theta}\pfrac{^2}{\varphi^2} \right)
\approx \frac{1}{r_s^2}\left(\pfrac{^2}{\theta^2} + \pfrac{^2}{\varphi^2}\right)
= \pfrac{^2}{\vz^2}\,,
\ee
%using $\sin\theta\approx 1$,
instead of the whole region ${0\leq\theta\leq\pi}$, $-\pi\leq\varphi\leq\pi$.
%\funai{approximation??}
{In this way,} we approximate the horizon sphere by a flat plane.
% which is not essential in our computation.
{This approximation can be justified, since our system has the spherical symmetry for $\theta, \varphi$ directions and, of course, any small region near the horizon is approximately flat spacetime.}

The equation (\ref{eq46}) %equation 
gives a common time $\tau$ to {both} inside and outside of the blackhole. 
Since the first term does not change by going 
from outside of the blackhole (real positive $\rho$) 
to inside the blackhole (imaginary negative $\rho$), 
we must change the sign of
$\mu^2 := \left(-\pfrac{^2}{\vz^2}+M^2\right)$
to keep the equation unchanged: {this means} % that is, 
$\mu$ must be imaginary.
\end{comment}

The Hamiltonian corresponding to the Rindler time $\tau$ can be commonly used both inside and outside of the blackhole.
This is the advantage of Rindler coordinate compared to the $(r, t)$ coordinates:
{in the latter case,} %in which case
we must exchange $r$ and $t$ when we go inside the blackhole, 
and also the Hamiltonian density changes from $\Theta_{00}$ to $\Theta_{rr}$.

By putting
%$\phi(\tau,\rho,\vz) =: e^{i(\vk\cdot\vz-\omega\tau)}\tphi(\rho)$,
$R_\ell(\tau,\rho) =: e^{-i\omega\tau}\tR_\ell(\rho)$,
we obtain
\be\label{eq48}
\left[ \left(\pfrac{}{\rho}\right)^2 + \frac{1}{\rho}\pfrac{}{\rho}
 - \left(\mu^2-\frac{\omega^2}{\rho^2} \right)\right] \tR_\ell=0
\ee
where 
$\mu^2 := \frac{\ell(\ell+1)}{r_s^2}+M^2$.
The solution of Eq.\,(\ref{eq48}) with energy $\omega$ %and transverse momentum $\vk$ 
is given by
\be\label{eq49}
\phi(\tau,\rho,\theta,\varphi)
= \frac{1}{N}{J}_{-i\omega}(i\mu\rho) e^{-i\omega\tau}Y_{\ell m}(\theta,\varphi)
=: K(\omega,\tfrac{\mu\rho}{2},\tfrac{\mu\rho}{2}) e^{-i\omega\tau}Y_{\ell m}(\theta,\varphi)
\ee
where
{$N$ is the normalization factor determined in Appendix \ref{sec:B}, and}
{$J$} % $Z$ 
is the Bessel (or Hankel) function with the integral formula
\be
\frac{1}{i\pi}K(\omega,\tfrac{\mu\rho}{2},\tfrac{\mu\rho}{2}) 
= \frac{1}{i\pi N} \int_C\frac{ds}{s}\,s^{-i\omega} e^{\frac{i\mu\rho}{2}(s-\frac{1}{s})}
\ee
where the path $C$ can be $[0, \infty]$.
\begin{comment}
and 
\be\label{eq51}
\mu^2=-\pfrac{^2}{\vz^2}+M^2 = \vk^2+M^2\,.
\ee
%,\qquad
%\vk=\left(\frac{2n_\theta}{\pi r_s}, \frac{n_\varphi}{\pi r_s}\right)\,,
%with $n_\theta, n_\varphi$ being integers.
{Let us discuss the condition for $\vk$. 
When we consider the whole region of $\theta\in [0,\pi]$ with $r=r_s$ fixed,
the scalar field $\phi(\tau,\rho,\vz)$ and $\mu^2$ are generalized as 
\be \label{eq51gen}
\phi(\tau,\rho,\theta,\varphi) = e^{-i\omega\tau}\tphi(\rho)Y_{\ell m}(\theta,\varphi)
\,,\qquad
\mu^2 = \frac{1}{r_s^2}\ell(\ell+1)+M^2\,,
\ee 
where the functions $Y_{\ell m}(\theta,\varphi)$ are the spherical harmonics with the integers $\ell=0,1,\ldots,\infty$ and $m=-\ell,-\ell+1,\ldots,\ell$.
Then we find the factor $e^{i\vk\cdot\vz}$ at $\theta\approx\frac{\pi}{2}$ 
must be written as a linear combination of $Y_{\ell m}$: 
% around $\theta\approx \frac{\pi}{2}$: 
\be \label{cond:k}
e^{ir_s\vk\cdot(\theta,\varphi)} \Big|_{\theta\approx\frac{\pi}{2}}
= \sum_{\ell, m} A_{\ell m} Y_{\ell m}(\theta\approx\tfrac{\pi}{2},\varphi)
\ee
where $\vk^2 = \frac{1}{r_s^2}\ell(\ell+1)$ and $A_{\ell m}$ are constant numbers. 
}

{Now we impose $\theta\approx \frac{\pi}{2}$ again.}

{Going from outside to inside of} the blackhole, we have
\be
\mu\rho\to -\mu\rho\,,\qquad
\vz \to i\vz\,.
\ee
By making $\vk\to -i \vk$, we have exactly the same equation as Eq.\,(\ref{eq48}).
\end{comment}
Then we note useful identities satisfied both by Bessel and Hankel functions:
\be
%K(\omega, \alpha, \beta) \to 
K(\omega, \alpha, \beta) = K^*(-\omega, -\alpha, -\beta).
\ee
For both Bessel and Hankel functions we also have
\be
K(\omega,\alpha,\beta) &=& e^{-\pi\omega}K^*(-\omega,\alpha,\beta)
\qquad(\text{for positive}~\alpha,\beta) \nt
K(\omega,\alpha,\beta) &=& e^{\pi\omega}K^*(-\omega,\alpha,\beta)
\qquad(\text{for negative}~\alpha,\beta).
\ee
In the end, we have to choose Bessel function so that the boundary value is finite.
Then we have
\be\label{eq54}
&&\phi(\tau,\rho,\theta,\varphi)
= \int_{-\infty}^{+\infty} d\omega \int\frac{d^2\vk}{\sqrt{2(2\pi)^{ 4}}} 
 K(\omega,\tfrac{\mu\rho}{2},\tfrac{\mu\rho}{2}) e^{-i\omega\tau}
 Y_{\ell m}(\theta,\varphi) 
 a(\omega,\ell,m) + h.c. \nt
&&= \int_0^{+\infty} d\omega \int\frac{d^2\vk}{\sqrt{2(2\pi)^{ 4}}} 
 K(\omega,\tfrac{\mu\rho}{2},\tfrac{\mu\rho}{2}) e^{-i\omega\tau} 
 Y_{\ell m}(\theta,\varphi) 
 \left[a_1(\omega,\ell,m) + a_2(\omega,\ell,m)\right]+ h.c. \nt
\ee
% [normalization is determined following 't Hooft's paper]
%From Eq.\,(\ref{eq54}) we obtain
where 
the region $\mu\rho < 0$ can be either 
inside of the blackhole or unphysical regions in the Penrose diagram. %[2] 
{$a(\omega,\ell,m)$ is the annihilation operator satisfying %(overall factor?)
\be
[a(\omega,\ell,m), a^*(\omega',\ell',m')] = \frac{1}{r_s^2}\delta(\omega-\omega')\delta_{\ell\ell'}\delta_{mm'}\,,\quad
\text{otherwise}=0\,,
\ee
and} we define
\be \label{bogolyu}
a_1(\omega,\ell,m)
&:=& \frac{1}{\sqrt{1-e^{-2\pi\omega}}} \left(
 a(\omega,\ell,m) + e^{-\pi\omega} (-1)^m a^*(-\omega,\ell,-m) \right), \nt
a_2(\omega,\ell,m)
&:=& \frac{1}{\sqrt{1-e^{{-}2\pi\omega}}} \left(
 e^{-\pi\omega} (-1)^m a^*(\omega,\ell,-m) + a(-\omega,\ell,m) \right),
\ee
{so that these operators satisfy
\be
&&[a_1(\omega,\ell,m),a_1^*(\omega',\ell',m')]=[a_2(\omega,\ell,m),a_2^*(\omega',\ell',m')]
=\frac{1}{r_s^2}\delta(\omega-\omega')\delta_{\ell\ell'}\delta_{mm'}\,,\nt
%=[a(\omega,\vk),a^*(\omega',\vk')]\,,\quad
&&\text{otherwise}=0\,. \nt
\ee
Note that $a_1(\omega,\vk)$ and $a_2(\omega,\vk)$ %these relations 
are defined only in the region $\omega\geq 0$.}
Solving these equations backwards, we obtain the relations
\be\label{eq56}
a_1(\omega,\ell,m) - e^{-\pi\omega} (-1)^m a_2^*(\omega,\ell,-m)
&=& \sqrt{1-e^{-2\pi\omega}}\,a(\omega,\ell,m)\,,\nt
{a_2(\omega,\ell,m) - e^{-\pi\omega} (-1)^m a_1^*(\omega,\ell,-m)} 
&=& {\sqrt{1-e^{-2\pi\omega}}\,a(-\omega,\ell,m)\,.}
\ee

\subsection{Entangled state solution and entanglement entropy}

The blackhole ground state (Hawking ground state) is defined as
\be
a(\omega,\ell,m) |\Omega\rangle = 0\,.
\ee
In terms of $a_1(\omega,\ell,m)$ and $a_2(\omega,\ell,m)$,
using Eq.\,(\ref{eq56})
we have %from
\be
\left[a_1(\omega,\ell,m) - e^{-\pi\omega} (-1)^m a_2^*(\omega,\ell,m)\right]
|\Omega\rangle = 0\,.
\ee
The solution is, as given by 't Hooft~\cite{ref2},
\be\label{eq59}
|\Omega\rangle 
= \frac{1}{N_\Omega} e^{\int_0^{+\infty} d\omega 
%\int {\frac{r_s^2 d^2\vk}{\sqrt{2(2\pi)^4}}}\, 
\sum_{\ell,m} r_s^2 (-1)^m\, 
e^{-\pi\omega} a_1^*(\omega,\ell,m) a_2^*(\omega,\ell,-m)}
|0\rangle_1 |0\rangle_2
\ee
{where $N_\Omega$ is the normalization factor.}
Starting from this expression, we can easily calculate the von Neumann (entanglement) entropy of the system:
\be\label{eq60}
S_{EE} = %\frac{\cA}{{4\pi}} 
 \int_0^{+\infty} d\omega \,\cK \left[
 -\log\left(1-e^{-2\pi\omega}\right)
 + \frac{2\pi\omega}{e^{2\pi\omega}-1}  \right]
\ee
where
%{$\cA$ is the area of the blackhole horizon,
%and  
we define 
$\cK:=\sum_{\ell,m} 1$
%$\cK:=\int \frac{d^2\vk}{\sqrt{2(2\pi)^4}}$}.
%$K:=\int d^2\vk$.
and it %This $\cK$ 
can be calculated as follows: 
The maximum value of the angular momentum $\ell$ at the horizon is given by, ignoring the mass of the scalar particle,
\be
\ell_{\rm max} = r_s \times \frac{\omega}{2r_s} = \frac{\omega}{2}
\ee
then we have
\be
\cK = \sum_{\ell,m} 1 = \sum_{\ell} (2\ell+1) = \ell_{\rm max}(\ell_{\rm max}+2) = \frac{\omega}{4} (\omega+4)\,.
\ee
The detail of this calculation is given in Appendix \ref{sec:A}.

\subsection{Hawking radiation}

We now calculate Hawking radiation from Eq.\,(\ref{eq39}):
\be \label{eq39'}
R_H 
= {-i} \langle\Omega| [H_A,H_B] |\Omega\rangle 
=  r_s^2\left(1-\frac{r_s}{r}\right)
   \int_{bdy} {\sin\theta} % \theta\approx \frac{\pi}{2} 
   d\theta d\varphi\, 
   \langle\Omega| \Theta_{0r} |\Omega\rangle \Big|_{r=r_s}\,.
\ee
We write down the result, leaving the details to Appendix \ref{sec:B}:
\be\label{eq61}
R_H = %{\sqrt{2} } %\frac{\cK}{8}
\frac{1}{4\pi r_s^2}
\int_0^{+\infty}d\omega\, \cK \frac{\omega}{e^{2\pi\omega}-1}
=
\frac{1}{16\pi r_s^2}
\int_0^{+\infty}d\omega\, \frac{\omega^2(\omega+4)}{e^{2\pi\omega}-1}\,.
\ee
This is similar to the black body radiation formula with the temperature $1/2\pi$.
However, one notices a slight deviation of this expression from the Planck formula for the black body radiation in which the integral must be
\be\label{eq62}
\sim \int_0^{+\infty}d\omega\, \frac{\omega^3}{e^{2\pi\omega}-1}\,.
\ee
Our radiation is not a thermal process and therefore we should be surprised by the similarity
of two formulae rather than their difference. 
\begin{comment}
We also note that in the equations
\be
H_B(t) = U^\dagger(t)H_B(0)U(t) = -i\int^t dt\,R_H
\ee
\be
i\pfrac{U}{t} = UH_A
\ee
\end{comment}

{To be more precise,}
we must take into account the back reaction of the radiation due to the decrease of the energy; {to say it simply}, % : {putting it simply,} %: most simply by taking into account 
the time dependence of the blackhole mass due to the radiation. 
In this sense, Eq.\,(\ref{eq61}) is just an approximation.

Another way to clarify the difference is that our radiation is from a pure (although entangled)
state
\be
|\Omega\rangle
= \frac{1}{N_\Omega} e^{\int_0^{+\infty} d\omega 
%\int {\frac{r_s^2d^2\vk}{\sqrt{2(2\pi)^4}}}\, 
\sum_{\ell,m}r_s^2 (-1)^m\,
   e^{-\pi\omega} a_1^*(\omega,\ell,m) a_2^*(\omega,\ell,-m)}
|0\rangle_1 |0\rangle_2\,.
\ee
On the other hand, the Planck radiation equation (\ref{eq62}) is from the mixed state 
{$|i\rangle$}
that can only be described by the density matrix $\hat\rho$:
\be
%\hat\rho = \sum_i e^{E_i/T} |i\rangle\langle i|
\hat\rho = \sum_i e^{2\pi\omega_i} |i\rangle\langle i|
\ee
%{where $E_i$ is energy of the state $|i\rangle$ and $T$ is temperature.}
We note that our density matrix for the entangled state which we obtain in Appendix \ref{sec:A}  
by summing up all the states inside the black hole is given by
\be
\hat\rho_1^N = 
 \sum_{\omega_i,\ell,m} \left(1-e^{-2\pi\omega_i}\right)^N 
 \sum_{n=0}^\infty e^{-2Nn\pi\omega_i} |n\rangle\langle n|\,.
\ee
This result means that, although we can treat the blackhole state as a mixed state by ignoring (or summing up) all the unknown (to the observers outside of the blackhole) states inside the blackhole, the density matrix is not the same as that of the constant temperature system. We see that,
%We can sum over states {of either $|~\rangle_1$ or $|~\rangle_2$} %in either 1 or 2 
%in Eq.\,(\ref{eq59}) and get a nontrivial density matrix, and then claim that we have a mixed state. 
%However, 
since the entire system is pure (coherent), the system utilizes the Hawing radiation (\ref{eq61}) to communicate and recover the coherence. 
The simplest example of EPR shows that the entanglement entropy has nothing to do with thermodynamics.
In this sense, we are claiming the EPR\,$=$\,SH (Steven Hawking) using the terminology
similar to EPR\,$=$\,ER~\cite{ref8}.

Our approach, at least in the spirit, is shared by some of the recent publications in which they claim that Hawking radiation carries information~\cite{ref9,ref9a,ref9b}\footnote{
The authors of these references discuss information loss in the context of AdS/CFT. 
They argue that the information inside a blackhole can be transferred to the Hawking radiation, while the semi-classical analysis around the horizon is kept intact. 
The information is not lost after the evaporation of a blackhole. 
Their system is a combination of classical (with Gibbons-Hawking entropy) and the quantum mechanical objects.}.

%%%%%%%%%%%%%%%%%%%%%%%%%%%%%%%%%%%%%%%%
%\section{Does a realistic blackhole evaporate?}
\section{Comments for realistic blackhole}
\label{sec:5}
%%%%%%%%%%%%%%%%%%%%%%%%%%%%%%%%%%%%%%%%

%\subsection{Hawking radiation and entanglement entropy}

We have already obtained Hawking radiation $R_H$ in Eq.\,(\ref{eq61})
and entanglemant entropy $S_{EE}$ in Eq.\,(\ref{eq60}).
%If we take the cutoff to be expressed in terms of the Planck scale, we have

With 
%\be
%\cK=\int \frac{d^2\vk}{\sqrt{2(2\pi)^4}} =: \frac{\xi}{G}
%\ee
$\cK=\sum_{\ell, m}1 = \frac14{\omega}(\omega+4)$,
%where $G$ is the Newton's constant.
the Hawking radiation (\ref{eq61}) becomes
\be
R_H
= \frac{1}{16\pi r_s^2} % {\sqrt{2}} \frac{\xi}{G} 
  \int_0^{+\infty}d\omega\, \frac{\omega^2(\omega+4)}{e^{2\pi\omega}-1}
= \frac{\pi^3 + 240 \zeta(3)}{3840 \pi^4 r_s^2}
%= \frac{\pi^5 + 45\zeta(5)}{960\pi^6 r_s^2}
%= {\frac{1}{12\sqrt{2}}} \frac{\xi}{G}
\ee
and the entanglement entropy (\ref{eq60}) becomes
\be
S_{EE} = %\frac{\cA}{{4\pi}} \frac{\xi}{G} % 2\pi^2 %r_s^2 K
 \int_0^{+\infty} d\omega \frac{\omega(\omega+4)}{4} \left[
 -\log\left(1-e^{-2\pi\omega}\right)
 + \frac{2\pi\omega}{e^{2\pi\omega}-1}  \right]
= \frac{\pi^3 + 270 \zeta(3)}{360 \pi^2}\,.
%= {\frac{1}{24}}\frac{\xi \cA}{G} % 12\pi
\ee
%where $\cA$ is the area of the blackhole horizon.
where $\zeta(s)$ is the Riemann zeta function.

%We note the following {comments on our results}:
%\begin{enumerate}
%\item 
%We have a cutoff independent 
Then we have a relation between the Hawking radiation and the entanglement entropy:
\be
S_{EE} = %{\frac{1}{\sqrt 2}} 
\frac{8 \pi}{3} \frac{\pi^3 + 270 \zeta(3)}{\pi^3 + 240 \zeta(3)}
%\frac{2 \pi^3}{3} \frac{\pi^3 + 270 \zeta(3)}{\pi^5 + 45 \zeta(5)}
\cA R_H\,.
%{\frac{16}{\pi}} A R_H\,.
\ee
Thus the entanglement entropy can be calculated in terms of observed total Hawking radiation $R_H$ and the surface area $\cA$ of the blackhole horizon.
This is consistent with that of Bekenstein-Hawking entropy.

{Finally, we comment on the fate of a realistic blackhole.}

It is customary to consider the Schwarzschild solution corresponding to a blackhole. It has
nothing inside the horizon except %for 
the singularity at the center. The information loss is
related to the existence of this singularity. If we {regard} %interpret 
a blackhole literally %regarding it
as %corresponds to 
the Schwarzschild solution, we cannot avoid the information loss
because the singularity at the single point cannot contain all the initial information.

However, in the previous sections,
we showed that the blackhole is not a mixed state and there must be no information loss.
{This strongly suggests that in our approach} we cannot take the Schwarzschild solution as it is. 
In fact, we can show that there exists a solution of the Einstein equation which has a horizon
outside of some hard solid core, {if we assume the density of matter has some finite upper bound, e.g., the Planck scale~\cite{tbp}.}
% (to be published).

Then, by assuming that there exists a solid core rather than a singularity inside the blackhole horizon, we can show that such a realistic blackhole actually does not evaporate.
{We will explain it in detail in a future work~\cite{tbp}.}

\section{Conclusion}
%%%%%%%%%%%%%%%%%%

In this paper, we calculate the Hawking radiation and the entanglement entropy of a blackhole based on the local field theory.
The Schwinger commutation relation which symbolized the locality of the quantum field theory is explicitly utilized.
%Both the Hawking radiation and the entanglement entropy turn out to be proportional to a common divergent quantity thus making the ratio to be finite.

The blackhole ground state is an entangled state of inside and outside,
and it makes sense to calculate the entanglement entropy.
However, it is nevertheless a pure state and the entropy is not that of a mixed state as is often claimed.
We also found that the formula we obtain for the Hawking radiation is similar to the black body radiation from a fixed temperature but not exactly,
so thus invalidating the interpretation of a blackhole with an object of fixed temperature.

As for the information loss, we do not have it since the local field theory is unitary. 
More concretely, our Hamiltonian of the quantum scalar field is Hermitian.
The inside region and the outside region of a blackhole communicate each other using the Hawking radiation.
This situation must be common to all the entangled states including the simple EPR state.

{Therefore, a blackhole never evaporates in our approach.}
We will show the realistic model of such a blackhole in a future work~\cite{tbp}:
it is the same as the Schwarzshild solution outside the horizon,
but inside it has a solid core to avoid the information loss.

\begin{comment}
Does a blackhole evaporate? Our answer is no.
A blackhole solution of the Einstein equation describes correctly a real blackhole near the horizon but not deep inside them.
The Hawking radiation makes the horizon radius smaller,
but after it hits the surface of the core material inside the blackhole,
and finally the blackhole turns back into a neutron star.
\end{comment}

\subsubsection*{Acknowledgment}

We thank Professor Jiro Arafune for his comments at early stage of this work.
We also thank Professor Alex Kusenko in UCLA for his valuable comments.

\appendix 

%%%%%%%%%%%%%%%%%%%%%%%%%%%%%%%%%%%%%%
\section{Calculation of entanglement entropy}
\label{sec:A}
%%%%%%%%%%%%%%%%%%%%%%%%%%%%%%%%%%%%%%

%Taking the discrete version of  with
\begin{comment}
{The integrals in Eq.\,(\ref{eq59}) are discretized as}
\be
\int d\omega\,\omega ~\to~ \sum_i \omega_i\,,\quad
\int \frac{r_s^2 d^2\vk}{\sqrt{2(2\pi)^4}} ~\to~
\sum_\vk\,.
%\frac{1}{4\sqrt{2} \pi r_s^2}\sum_{\ell=0}^\infty \sum_{m=-\ell}^{\ell}\,.}
%\frac{2}{\pi^2 r_s^2}\sum_{n_\theta,n_\varphi}
%=: \frac{2}{\pi^2 r_s^2}\sum_{\vk} \,,}
\ee
%{where the exact definition of $\sum_\vk$ will be discussed later.}
%since $\vk = \left(\frac{2n_\theta}{\pi r_s}, \frac{n_\varphi}{\pi r_s}\right)$
%and $n_\theta, n_\varphi$ are integers,
%as mentioned in Eq.\,(\ref{eq51}).
Then 
\end{comment}
First we obtain the discrete version of Eq.\,(\ref{eq59}):
\be \label{eq:A1}
|\Omega\rangle 
&=& \frac{1}{N_\Omega}
%\prod_{\omega_i,\vk} e^{e^{-\pi\omega_i} a_1^*(\omega_i,\vk) a_2^*(\omega_i,\vk)}
\prod_{\omega_i,\ell,m} e^{r_s^2(-1)^m e^{-\pi\omega_i} a_1^*(\omega_i,\ell,m) a_2^*(\omega_i,\ell,-m)}
|0\rangle_1 |0\rangle_2 \nt
&=& 
 \prod_{\omega_i,m} \sqrt{1-e^{-2\pi\omega_i}}
 \sum_{n=0}^\infty (-1)^{mn}e^{-n\pi\omega_i} |n\rangle_1 |n\rangle_2 \,.
\ee
The total density matrix is
\be
\hat\rho
&=& 
 \prod_{\omega_i,\ell,m} \sqrt{1-e^{-2\pi\omega_i}}
 \sum_{n=0}^\infty (-1)^{mn}e^{-n\pi\omega_i}
 \prod_{\omega_i',\ell',m'} \sqrt{1-e^{-2\pi\omega_i'}}
 \sum_{n'=0}^\infty (-1)^{m'n'}e^{-n'\pi\omega_i'}\nt
&&\times 
  |n\rangle_1 |n\rangle_2 \langle n'|_1 \langle n'|_2 \,.
\ee
Taking the trace {for the states $|n\rangle_2$}, 
we get the density matrix for the system 1.
{In addition,} we apply the replica method here:
\be\label{rho}
\hat\rho_1^N = 
 \sum_{\omega_i,\ell,m} \left(1-e^{-2\pi\omega_i}\right)^N 
 \sum_{n=0}^\infty e^{-2Nn\pi\omega_i} |n\rangle\langle n|\,.
\ee
Then the entanglement entropy is calculated as
\be
S_{EE} 
&=& -{\rm Tr}\, \rho_1\log\rho_1
 = -{\rm Tr}\, \pfrac{\rho_1^N}{N}\bigg|_{N=1} \nt
&=& -{\rm Tr}\, \pfrac{}{N}
 \prod_{\omega_i,\ell,m} \sum_{n=0}^\infty 
 \left[\left(1-e^{-2\pi\omega_i}\right) e^{-2n\pi\omega_i}\right]^N\bigg|_{N=1}
 |n\rangle\langle n| \nt
%&=& - \pfrac{}{N}
% \prod_{\omega_i} \sum_{n=0}^\infty 
% \left[\left(1-e^{-2\pi\omega_i}\right) e^{-2n\pi\omega_i}\right]^N\bigg|_{N=1} \nt
&=& - \pfrac{}{N}
 \prod_{\omega_i,\ell,m}
  \frac{\left(1-e^{-2\pi\omega_i}\right)^N}{1-e^{-2Nn\pi\omega_i}} \bigg|_{N=1}\,.
\ee
\begin{comment}
%In the continuous limit, it becomes
{Let us now take the continuous limit. 
Here we note that the integrated function does not depend on $\vk$:
this means $\sum_\vk$ just counts the number of $\vk$, up to a constant factor.
The condition for $\vk$ was already shown in Eq.\,(\ref{cond:k}),
then we find 
\be
\sum_\vk \propto \sum_{\ell=0}^\infty \sum_{m=-\ell}^{\ell}
~\to~ \int_0^\infty d\ell \int_{-\ell-\frac12}^{\ell+\frac12} dm
\ee
if we consider the continuous space of $\ell, m$ (the indices of $Y_{\ell m}$). 
On the other hand, 
using the relation $|\vk|^2 = \frac{1}{r_s^2}\ell(\ell+1)$, 
we obtain
\be
\int d^2\vk 
= \int_0^\infty d|\vk| \cdot |\vk|\int_{-\pi}^{+\pi} d\theta_k
= \frac{\pi}{r_s^2}\int d\ell dm\,.
%~\to~ \frac{\pi}{r_s^2}\sum_{\ell,m}
\ee
Here we can simply set $|\vk|d\theta_k = \frac{2\pi|\vk|}{2\ell+1} dm$,
since the integrated function is independent of $\vk$.
Then the sum $\sum_\vk$ here can be written as
\be
\sum_\vk ~\to~ \frac{r_s^2}{\pi}\int d^2\vk
\ee
} 
\end{comment}
In the continuous limit, and we obtain
%\sum_{\ell,m} = \sum_{\ell=0}^\infty \sum_{m=-\ell}^{\ell} =: \sum_{\vk}\,.
\be
S_{EE} 
&=& - \pfrac{}{N}
 e^{\int d\omega \sum_{\ell,m} \,\log
 \frac{\left(1-e^{-2\pi\omega}\right)^N}{1-e^{-2Nn\pi\omega}}} \bigg|_{N=1} \nt
&=& - \pfrac{}{N}
  \int d\omega \sum_{\ell,m} %\int \frac{r_s^2d^2\vk}{\sqrt{2(2\pi)^4}}
  \,\log
  \frac{\left(1-e^{-2\pi\omega}\right)^N}{1-e^{-2Nn\pi\omega}} \bigg|_{N=1} \nt
&=& 
  \int d\omega \sum_{\ell,m} %\int \frac{r_s^2d^2\vk}{\sqrt{2(2\pi)^4}} 
  \left[
  -\log\left(1-e^{-2\pi\omega}\right)
  +\frac{2\pi\omega}{e^{2\pi\omega}-1}\right]\,.
\ee
%\funai{$d^2\vk = r_s^2\sum_\vk$ (??)}
Finally, we define $\cK:=\sum_{\ell,m}$ % \int \frac{d^2\vk}{\sqrt{2(2\pi)^4}}$} 
and obtain Eq.\,(\ref{eq60}):
%\be
%S_{EE} 
%&=& {\frac{\pi^2 r_s^2 \cK}{2}} \int d\omega \left[
%  -\log\left(1-e^{-2\pi\omega}\right)
%  +\frac{2\pi\omega}{e^{2\pi\omega}-1}\right]\,.
%\ee
\be
S_{EE} 
&=& %{\frac{\cA}{4\pi}}  
  \int d\omega\, \cK \left[
  -\log\left(1-e^{-2\pi\omega}\right)
  +\frac{2\pi\omega}{e^{2\pi\omega}-1}\right]\,.
\ee
%{with the surface area $\cA=4\pi r_s^2$ for the sphere with radius $r=r_s$.}
%\funai{careful calculation is required: flat space $A=2\pi^2 r_s^2$ or sphere $A=4\pi r_s^2$.}

%%%%%%%%%%%%%%%%%%%%%%%%%%%%%%%%%%
\section{Calculation of Hawking radiation}
\label{sec:B}
%%%%%%%%%%%%%%%%%%%%%%%%%%%%%%%%%%

{In our definition, the Hawking radiation is given as Eq.\,(\ref{eq39'}):}
\be
R_H 
&=& \frac{\rho^2}4 % r_s^2\left(1-\frac{r_s}{r}\right)
   \int_{bdy} {\sin\theta} 
   d\theta d\varphi\,
   \langle\Omega| \Theta_{0r}(r,\theta,\varphi) |\Omega\rangle \Big|_{r\approx r_s} %\nt
%&=& {4\pi} r_s^2\left(1-\frac{r_s}{r}\right)
   %\int%_{\theta\approx \frac{\pi}{2}} %\int_{bdy} 
   %d\varphi\, % d\theta 
   %\langle\Omega| \Theta_{0r} |\Omega\rangle \Big|_{r=r_s, \theta\approx \frac{\pi}{2}}\,.
\ee
where $\rho = 2r_s\sqrt{\frac{r}{r_s}-1}$.

The fact that $\langle\Omega| \Theta_{0r} |\Omega\rangle \neq 0$
means, in analogy to the language  of spontaneous breaking,
the Hawking ground state $|\Omega\rangle$
corresponds to the ``spontaneously broken" Lorentz invariant vacuum with the order parameter
$\langle\Omega| \Theta_{0r} |\Omega\rangle$.
The Bogolyubov transformation defined in Eq.\,(\ref{bogolyu}) % section 3 
can be interepreted along this line of thought. This shows that the vacuum
$|0\rangle_1\otimes |0\rangle_2$
corresponds to the ``symmetric" vacuum with
$\langle0|_1\langle0|_2\Omega \Theta_{0r} |0\rangle_1|0\rangle_2 = 0$.

Since the blackhole is a finite system,  there is no actual spontaneous breaking and the state
$|\Omega\rangle$
is in fact written down in terms of $|n\rangle_1\otimes |n\rangle_2$
as discussed in %Sec.\,\ref{sec:3.3} and 
Appendix \ref{sec:A}.

%due to the rotational symmetry, 
%we only consider the region of $\sin\theta\approx 1$ in the following calculations.
%Note that the factor $4\pi$ comes from $\int\sin\theta d\theta d\varphi = 4\pi$.
%First 
Then we have, using Eq.\,(\ref{eq54}), % and (\ref{eq39'}), 
\be\label{eqB2}
\Theta_{0r}\Big|_{r\approx r_s} % , \theta\approx \frac{\pi}{2}} % (r,\theta,\varphi) 
&=& \pfrac{\tau}{t}\pfrac{\rho}{r} \left(-\partial_\tau \phi \partial_\rho \phi \right) 
%= -\frac{\partial_\tau \phi \partial_\rho \phi}{2r_s\sqrt{\frac{r}{r_s}-1}}
\nt
&=&
\frac{1}{\rho} % {2r_s\sqrt{\frac{r}{r_s}-1}}
\left( \int_{-\infty}^{+\infty} d\omega \sum_{\ell,m} %\int\frac{d^2\vk}{\sqrt{2(2\pi)^{ 4}}} 
 i\omega
 K(\omega,\tfrac{\mu\rho}{2},\tfrac{\mu\rho}{2}) Y_{\ell m} e^{-i\omega\tau} %(\vk\cdot\vz 
 a(\omega,\ell,m) + h.c.\right) \nt 
&&\times
\left( \int_{-\infty}^{+\infty} d\omega' \sum_{\ell',m'} % \int\frac{d^2\vk'}{\sqrt{2(2\pi)^{ 4}}} 
 \partial_\rho K(\omega',\tfrac{\mu\rho}{2},\tfrac{\mu\rho}{2}) Y_{\ell'm'} e^{-i\omega'\tau} 
 a(\omega',\ell',m') + h.c.\right)\,.
\ee

The fact that $|0\rangle_{1,2}$ 
satisfies $\langle0|\Omega \Theta_{0r} |0\rangle = 0$
shows that we must define the normal ordering in terms of $a_{1,2}$ rather than $a$ itself 
in the above definition of $\Theta_{0r}(r,\theta,\varphi)$.
Then we use the expression (\ref{eq:A1}),
rewrite $\Theta_{0r}$ in terms of $a_1$ and $a_1^*$ rather than $a$ and $a^*$,
and we obtain
\be\label{W}
\langle\Omega| \Theta_{0r} |\Omega\rangle
&=& \prod_{\omega_i,\ell,m} \left(1-e^{-2\pi\omega_i}\right)
 \sum_{n=0}^\infty e^{-2n\pi\omega_i}
 \langle n|_1 \Theta_{0r}(r,\theta,\varphi)|n\rangle_1 \nt
&=&
{\rm Tr}\left[ \Theta_{0r}(r,\theta,\varphi)\hat\rho_1\right]
\ee
where
$\hat\rho_1$ is defined in Eq.\,(\ref{rho}) and 
\be
\Theta_{0r}%\Big|_{r\approx r_s} % , \theta\approx \frac{\pi}{2}} % (r,\theta,\varphi) 
&=&
\frac{1}{\rho} % {2r_s\sqrt{\frac{r}{r_s}-1}}
\left( \int_{-\infty}^{+\infty} d\omega \sum_{\ell,m} %\int\frac{d^2\vk}{\sqrt{2(2\pi)^{ 4}}} 
 i\omega
 K(\omega,\tfrac{\mu\rho}{2},\tfrac{\mu\rho}{2}) Y_{\ell m} e^{-i\omega\tau} %(\vk\cdot\vz 
 \sqrt{1-e^{-2\pi\omega}}a_1(\omega,\ell,m) + h.c.\right) \nt 
&&\times
\left( \int_{-\infty}^{+\infty} d\omega' \sum_{\ell',m'} % \int\frac{d^2\vk'}{\sqrt{2(2\pi)^{ 4}}} 
 \partial_\rho K(\omega',\tfrac{\mu\rho}{2},\tfrac{\mu\rho}{2}) Y_{\ell'm'} e^{-i\omega'\tau} 
 \sqrt{1-e^{-2\pi\omega'}}a_1(\omega',\ell',m') + h.c.\right)\,.\nt
\ee
This shows that an observer on the horizon observes the same Hawking radiation as an outside observer. 
We use Eq.\,(\ref{W}) to calculate the Hawking radiation.

Outside of the horizon, we have $\phi$ in terms of $a_1(\omega,\ell,m)$:
\be \label{out}
\phi = \int_0^{+\infty}d\omega \sum_{\ell,m}
 K(\omega,\tfrac{\mu\rho}{2},\tfrac{\mu\rho}{2}) Y_{\ell m} e^{-i\omega\tau} %(\vk\cdot\vz 
 \sqrt{1-e^{-2\pi\omega}}a_1(\omega,\ell,m) + h.c.
\ee
Then
%Collecting these contributions %in equations (B7) and (B9) 
%and substituting them into Eq.\,(\ref{eqB2}), 
we obtain
\be
R_H
&=& -i{\rm Tr}\left([H_A,H_B]\hat\rho_1\right) \nt
&=& \frac{\rho^2}{4}\int_{bdy} \sin\theta d\theta d\varphi\,
{\rm Tr}\left[\Theta_{0r}(r,\theta,\varphi)\hat\rho_1\right]\nt
&=&
\frac{\rho}{4r_s^2}
\int_0^{+\infty} d\omega \sum_{\ell,m}
i\omega K(\omega,\tfrac{\mu\rho}{2},\tfrac{\mu\rho}{2}) 
\partial_\rho K(\omega,\tfrac{\mu\rho}{2},\tfrac{\mu\rho}{2})
e^{-2\pi\omega} + c.c.
\ee  
%&=& {\pi} \rho^2
%  \langle\Omega| \Theta_{0r} |\Omega\rangle \Big|_{r=r_s,\theta\approx \frac{\pi}{2}}\nt
%&=& {-2\pi i} \rho 
%  \int_{0}^{+\infty} d\omega \int\frac{d^2\vk}{{2(2\pi)^{ 4}}} 
%  \omega
%  K(\omega,\tfrac{\mu\rho}{2},\tfrac{\mu\rho}{2}) 
 % \partial_\rho K^*(\omega,\tfrac{\mu\rho}{2},\tfrac{\mu\rho}{2}) 
  %e^{i(\vk\cdot\vz-\omega\tau)} 
  %a(\omega,\vk) + h.c.
%  \frac{1}{e^{2\pi\omega}-1} \bigg|_{r=r_s,\theta\approx \frac{\pi}{2}} \nt
  %\frac{e^{-2\pi\omega}}{1-e^{-2\pi\omega}} \nt
%&&
%  +\,c.c.
%Here we note that $r\to r_s$ is equivalent with 
in the limit of $\rho\to 0$.
Here we used
\be
\int\sin\theta d\theta d\varphi\, Y_{\ell m}^* Y_{\ell m} = 1\,.
\ee
%We calculate $K(\omega, \frac{\mu\rho}{2}, \frac{\mu\rho}{2})$ 
Using the expression of Bessel function (\ref{eq49}),
\be\label{eqB16}
\lim_{\rho\to 0} \rho
  K(\omega,\tfrac{\mu\rho}{2},\tfrac{\mu\rho}{2}) 
  \partial_\rho K^*(\omega,\tfrac{\mu\rho}{2},\tfrac{\mu\rho}{2})  
&=&
\frac{1}{N^2}
\lim_{\rho\to 0} \rho
  J_{i\omega}(i\mu\rho)
  \partial_\rho J^*_{i\omega}(i\mu\rho) \nt
&=&
 -\frac{i\omega e^{-\pi\omega}}{N^2}\,,
\ee
where we use the formula for the Bessel function on the boundary:
\be \label{eqB15}
\lim_{\rho\to 0}J_{i\omega}(i\mu\rho) = \left(\frac{i\mu\rho}{2}\right)^{i\omega}\,,\quad
\lim_{\rho\to 0}\partial_\rho J_{i\omega}^*(i\mu\rho) = \frac{-i\omega}{\rho}\left(\frac{-i\mu\rho}{2}\right)^{-i\omega}\,.
\ee
Therefore, we obtain
\be
R_H = 
%{-}\frac{\cK}{{\sqrt{2}\pi}} \int d\omega\, \frac{e^{\pm\pi\omega}}{N^2}
%  \frac{\omega^2}{e^{2\pi\omega}-1}\,,
\frac{1}{2r_s^2} \int_0^{+\infty}d\omega\, \cK\, \frac{\omega^2 e^{-3\pi\omega}}{N^2}
\ee
where %{$\cK:=\int \frac{d^2\vk}{\sqrt{2(2\pi)^4}}$} 
$\cK = \sum_{\ell, m} 1$ as before
and the value of $N$ is calculated below.

To determine the normalization factor $N$, we use the commutation relation (\ref{eq28}):
\be
[\partial_0\phi, \phi]
&=& -\frac{i}{g^{00}}\delta^3(\vx-\vx')
= \frac{ir_s}{r}\left(\frac{r}{r_s}-1\right)\delta(r-r')%\delta^2(\vz-\vz')
 \delta(\cos\theta-\cos\theta')\delta(\varphi-\varphi') \nt
&=& -i\frac{\rho^2}{4r_s^4} \delta\left(\frac{\rho^2}{4r_s} - \frac{\rho'^2}{4r_s}\right)
 \delta(\cos\theta-\cos\theta')\delta(\varphi-\varphi')
\ee
%where $\vz = r_s\times(\theta,\varphi)|_{r=r_s, \theta\approx\frac{\pi}{2}}$,
and %at the boundary 
we obtain
\be\label{eqB19}
[\partial_\tau\phi,\phi]
=\pfrac{t}{\tau} [\partial_0\phi, \phi]
=-i\frac{\rho}{r_s^2}\delta(\rho-\rho') \delta(\cos\theta-\cos\theta') \delta(\varphi-\varphi')\,.
\ee
On the other hand, 
using Eq.\,(\ref{out}) % eq54})
and the Bessel function formula (\ref{eqB15}), 
we obtain
\be\label{eqB20}
[\partial_\tau\phi,\phi]
&=& 
-\frac{1}{r_s^2}
\int_0^{+\infty}d\omega\, i\omega
  K(\omega,\tfrac{\mu\rho}{2},\tfrac{\mu\rho}{2}) 
  K^*(\omega,\tfrac{\mu\rho'}{2},\tfrac{\mu\rho'}{2})
  \left(1-e^{-2\pi\omega}\right)\nt
&&
  \times \delta(\cos\theta-\cos\theta')\delta(\varphi-\varphi')
 + h.c. \nt
&=&
-\frac{i}{r_s^2} \delta(\log\rho-\log\rho')
  \delta(\cos\theta-\cos\theta')\delta(\varphi-\varphi')
\ee
\begin{comment}
  -\frac{i}{4\pi}
  \int_{-\infty}^{+\infty} \frac{d\omega}{2\pi}\,
  %\int \frac{d^2\vk}{2(2\pi)^4} 
  \omega  
  K(\omega,\tfrac{\mu\rho}{2},\tfrac{\mu\rho}{2}) 
  K^*(\omega,\tfrac{\mu\rho'}{2},\tfrac{\mu\rho'}{2}) 
  \delta^2(\vz-\vz')
  + h.c. \nt
&\stackrel{\rho\to 0}{=}& 
  -\frac{i}{4\pi}
  \int_{-\infty}^{+\infty} \frac{d\omega}{2\pi}\,
  \frac{\omega e^{\pm \pi\omega}}{N^2}
  e^{i\omega(\log\rho-\log\rho')}
  \delta^2(\vz-\vz')
  + h.c.
\end{comment}
where we used 
\be
\sum_{\ell,m} Y_{\ell m}Y^*_{\ell m} = \delta(\cos\theta-\cos\theta') \delta(\varphi-\varphi')\,.
\ee
%In the limit $\rho\to 0$ (or $r\to r_s$) 
Then we find that Eqs.\,(\ref{eqB19}) and (\ref{eqB20}) are coincident when
\be
N^2 = 2\pi\omega e^{-\pi\omega}\left(1-e^{-2\pi\omega}\right)
%-\frac{\omega e^{\pm \pi\omega}}{{2}\pi}
\ee
is satisfied (for $\omega>0$). Finally, we obtain
\be
R_H = \frac{1}{4\pi r_s^2}  \int_0^{+\infty} d\omega\, \cK
\frac{\omega}{e^{2\pi\omega}-1} \,.
\ee

\end{document}